\documentclass[11pt,prd,preprintnumbers,eqsecnum,superscriptaddress,nofootinbib]{revtex4}
\usepackage[usenames,dvipsnames]{xcolor}
\usepackage{graphics,epsfig}
\usepackage{graphicx}
\usepackage{color}
\usepackage{amssymb,amsmath,amsfonts}
 \usepackage{amsmath}
\usepackage{amssymb}
\usepackage{hyperref}
\usepackage{epstopdf}

\def\td#1{\tilde{#1}}
\def\check{ \maltese {\bf Check!}}

\begin{document}
\def\td#1{\tilde{#1}}
\def\check{ \maltese {\bf Check!}}

	\vspace{25mm}
	
	\begin{center}
		{\Large \bf Generalized parton distributions of a deuteron in an AdS/QCD hard-wall model}
		
		\vskip 1. cm
	
		{Minaya Allahverdiyeva $^{a,c,d}$\footnote{e-mail: minaallahverdiyeva@ymail.com},
             Shahin Mamedov $^{a,b,c}$\footnote{e-mail: sh.mamedov62@gmail.com}}
		
		\vskip 0.5cm
		{\it $^a\,$ Institute of Physics, Ministry of Science and Education, H.Javid 33, Baku, AZ 1143, Azerbaijan}\\
		{\it $^b\,$ Institute for Physical Problems, Baku State University, Z.Khalilov 23, Baku, AZ 1148, Azerbaijan}\\
            {\it $^c\,$ Center for Theoretical Physics, Khazar University, 41 Mehseti Str., Baku, AZ1096, Azerbaijan} \\
              {\it $^d\,$ Western Caspian University, 31 Istiqlaliyyat Str., Baku, AZ1001, Azerbaijan}\\
        
	\end{center}

\centerline{\bf Abstract} \vskip 4mm

We investigate the gravitational form factors (GFFs) and the generalized parton distributions (GPDs) of the deuteron within the framework of the hard-wall AdS/QCD model. The momentum dependence of the GFFs obtained here is in good agreement with the results of the soft-wall AdS/QCD model. The value of the gravitational mean square radius in this model agrees with the experimental data. GFFs provide a holographic description of the. Deuteron GPDs are obtained from GFFs via sum rules and have shapes of plots similar to those for GPDs extracted from the electromagnetic form factors (EFFs) calculated in the framework of the soft-wall AdS/QCD  model. 

\vspace{1cm}
    \section{Introduction}

Understanding the internal dynamics of the hadrons is a central goal of modern nuclear and particle physics. In particular, the deuteron, as the simplest nuclear-bound state of a proton and a neutron, provides a unique testing ground for exploring the interplay between QCD dynamics and nuclear forces. Beyond electromagnetic form factors, the study of the deuteron through its energy-momentum tensor (EMT) opens access to knowledge about fundamental quantities such as the distribution of energy, momentum, and forces inside the nucleus~\cite{Muller, Ji3, Ji, Pol2}.

The elements of the matrix of the EMT are parametrized in terms of GFFs, which encode fundamental information on the mass, spin, and mechanical properties of hadrons and nuclei \cite{Bro, Pol, KMSh}. Through exact QCD sum rules, the GFFs are directly related to the GPDs, making them experimentally accessible in hard exclusive processes such as deep virtual Compton scattering (DVCS) and exclusive meson production~\cite{4}. 

The tensor structure of the deuteron offers valuable insights into the distributions and dynamics of quarks and gluons within the nucleus. It can be investigated experimentally through inclusive and semi-inclusive Deep Inelastic Scattering (DIS) of electrons on tensor-polarized deuteron targets. In inclusive DIS, one-dimensional (longitudinal momentum–dependent) tensor structure functions are extracted, while semi-inclusive DIS provides access to three-dimensional tensor structure functions that also depend on transverse momentum. Although experimental studies have long been pursued in this field \cite{expA, expC, expF}, achieving sufficiently high tensor polarization remains a significant challenge. In recent years, considerable progress has been made in improving the tensor polarization of deuteron targets \cite{Compass, Alice}. These advances open new opportunities for more precise and comprehensive investigations of the deuteron tensor structure.

The extraction of GPDs from experimental data remains inherently model dependent, as it requires phenomenological parametrizations of their functional form and underlying kinematic behavior~\cite{Aslan, Dima}.

Direct experimental access to gravitational form factors is not presently possible because gravity is exceedingly weak at hadronic length scales. However, their physical content can still be probed through their exact relation to GPDs, which provide a multidimensional description of the hadron and nuclear structure. The GPDs are obtained from the measurements in hard exclusive reactions, including deeply virtual Compton scattering~\cite{Bel} and deeply virtual meson production, where specific moments of the cross sections are directly linked to the underlying gravitational form factors. In parallel with experimental studies, the GFFs have been explored in numerous theoretical frameworks designed to describe nonperturbative QCD dynamics, such as lattice QCD~\cite{Lat}, QCD sum rule techniques~\cite{Ber}, chiral effective theories, topological soliton models, light-front quark models, and related phenomenological approaches. In this way, deuteron GPDs provide a crucial bridge between QCD at the quark–gluon level and the macroscopic mechanical properties of the nucleus encoded in the EMT.

Among the observables derived from GFFs, the gravitational radius plays a role analogous to the charge radius, characterizing the spatial distribution of the deuteron’s mass and energy. 

Since QCD is strongly coupled at the scales relevant for nuclear binding, direct perturbative methods are not applicable. Holographic models based on gauge/gravity duality (\cite{Mal, Gubs, Wit}) provide an alternative framework to investigate nonperturbative phenomena. In particular, the hard-wall AdS/QCD model (\cite{Boschi1, Boschi2, Erl}) implements confinement by introducing an infrared cutoff in AdS$_5$ space. This model presents numerous opportunities for the analytical calculation of hadronic form factors and related quantities.  The investigation of GFFs within the holographic hard-wall model was initiated with the works \cite{Abidinr, 3, AbN}. Next, this approach was extended to the soft-wall model and applied to the investigation of GPDs \cite{ Lubsoft, Sharma, Mro}. The holographic study of the deuteron form factors was done in the next works \cite{lyudeyt, lyudeyt2, 6, Mondal, Md, Shr}. Also, the GFFs are studied in the top-down and light front approaches of holographic QCD \cite{Fujii, Fujii2, Fujii3, Brod}. 

In this work, we compute the deuteron GFFs and establish their relationship to GPDs within the hard-wall AdS/QCD model. From these results, we extract the deuteron gravitational radius and compare our findings with lattice QCD, phenomenological parameterization, and available experimental constraints.

This work is organized as follows. In Section II, we present the basic elements of the hard-wall model. In Section III, study the deuteron GFFs and gravitational radius. In Section IV, we study deuteron GPDs obtained from the GFFs. Summary and conclusions are presented in Section V.

\section{Deuteron and graviton in hard-wall model}

In the AdS / QCD hard-wall approach, the deuteron as a spin-1 particle is described by the vector field  (twist 6) propagating in the five-dimensional AdS space with the infrared cutoff at $z=z_0$. The background geometry is perturbed by the $h_{\mu\nu}$ field  \cite{Abidinr,Mro},  
\begin{equation}
ds^2=g_{MN}dx^Mdx^N=\frac{1}{z^2}\left((\eta_{\mu\nu}+h_{\mu\nu})dx^\mu dx^\nu - dz^2\right), \qquad 0<z\leq z_0, 
\end{equation}
where $\eta_{\mu\nu}=\mathrm{diag}(1,-1,-1,-1)$ is the flat Minkowski metric, the wall $z=\epsilon$, understood by $\epsilon \rightarrow 0$, corrosponds to the UV limit of QCD, and the wall located at $z=z_0$ sets the scale for the breaking of the conformal symmetry of QCD in the IR region and $h_{\mu\nu}$ in the Randall-Sundrum gauge \cite{Jaume}, where $h_{\mu\nu}$ is transverse and traceless (TT) part of the Minkowski metric perturbation. Also, it satisfies the axial gauge condition $h_{z\mu} = h_{zz} = 0$. The Greek indices will run from $0$ to $3$, and the Latin indices will run from $0$ to $5$.

The dynamics of the bulk deuteron field $d_M(x,z)$ is governed by the five-dimensional action \cite{lyudeyt, Md},
\begin{equation}
S=\int d^5x \, \sqrt{g} \, \Big[ -\partial ^M d^\dagger_{N} \partial_M d^N 
+ d^\dagger_{M} \, \mu^2  \, d^M + \mathcal{R} + 12 \Big],
\end{equation}
where $\mathcal{R}$ is the Ricci scaliar of AdS$_5$.  

The bulk mass $\mu^2$ is related to the conformal dimension $\Delta$ of the deuteron interpolating operator in the boundary theory via the AdS/CFT relation,  
\begin{equation}
\mu^{2} = (\Delta-1)(\Delta-3) = (L+5)(L+3),
\end{equation}
where $L$ denotes the quantum number of orbital angular momentum and $\Delta=L+\tau$ ($\tau=6$ is the twist number). For the ground state of deuteron $L=0$ and this yields $\Delta=6$.

\subsection{Deuteron Profile Function in the Hard-Wall AdS/QCD model}

In the axial gauge $d_z=0$, the deuteron bulk field is decomposed as~\cite{lyudeyt, lyudeyt2,6}
\begin{equation}
d_\mu(x,z)=\int \frac{d^4q}{(2\pi)^4}\, e^{-iq\cdot x}\, d_\mu(q)\,\psi(q,z),
\end{equation}
where $\psi(q,z)$ denotes the holographic profile of the deuteron along the fifth dimension. Using the Kaluza-Klein (KK) decomposition of the vector AdS field dual to the deuteron states
\begin{equation}
    d^\mu(x,z)=\exp{\Big[-\frac{A(z)}{2}\Big]} \sum_n d^\mu_n(x)\psi_n(z).
\end{equation}

The quadratic part of the action for the vector field leads to the equation of motion~\cite{6}
\begin{equation}\label{eq:eom}
\Big[-\partial_z^2 + \frac{4(L+4)^2-1}{4z^2}\Big]\psi_n(z)=m^2_n\psi_n(z),
\end{equation}
which is equivalent to a Bessel equation of order four. $q^2$ was replaced by the Kaluza-Klein modes mass spectrum $m^2_n$, which corresponds to the deuteron spectrum. The general solution for the $L=0$ state has the following form:
\begin{equation}\label{eq:profile}
\psi_n(z)=b_1 \sqrt{z}\, J_4(m_n z) + b_2 \sqrt{z}\, Y_4(m_n z).
\end{equation}
The regularity at $z\to 0$ requires $b_2=0$, leaving only the solution $J_4$.  

The boundary conditions for this solution are the following:
\begin{eqnarray}
    \partial_z \psi_n(z)\big|_{z=z_0}=0, \nonumber\\
\psi_n(z=0)=0. \label{2.8}
\end{eqnarray}
 Normalized solutions
\begin{equation}
\psi_n(z)=\frac{\sqrt{z}\,J_4(m_n z)}{\sqrt{\int^{z_0}_0 dz \ \left[J_4(m_n z)\right]^2}},
\end{equation}
satisfy the normalization condition $\int (dz/z)\psi^2_n(z)=1$.  Using solution (\ref{eq:profile}) and boundary conditions (\ref{2.8}), repeating the procedure to fix $z_0$ developed in \cite{motoi}, we find the following relation between $z_0$ and $m_n$ :
\begin{equation}
    -\frac{1}{z_0}=-2m_n\tan\left(m_nz_0-\frac{9\pi}{4}\right).
\end{equation}
This relation gives us a quantized spectrum for the Kaluza-Klein modes:
\begin{equation}
    m_n\simeq\left(n+\frac{9}{4}\right)\pi z_0^{-1} ~\quad (n=1,2,...).
\end{equation}
Denoting the mass of the ground state ($n=1$) of the deuteron by $m_D$, we get a relation between this mass and the $z_0$ parameter of the model: 
\begin{equation}
    m_D\simeq\frac{13}{4}\pi z^{-1}_0=10.2z^{-1}_0. \label{2.13}
\end{equation}
In the ground state, the deuteron mass known from experiment is $m_D=1.876~\mathrm{GeV}$, and we fix $z_0$ by this mass using (\ref{2.13}):
\begin{equation}
   z_0=5.44 ~\rm{GeV^{-1}}.
\end{equation}

\subsection{Bulk-to-Boundary Propagator of Graviton}

In AdS/QCD correspondence theory, the binding of the deuteron to the QCD EMT occurs through its interaction with the bulk graviton field. The response of the AdS geometry to the external graviton source is described by the graviton bulk-to-boundary propagator, which encodes how a perturbation injected at the UV boundary propagates into the bulk.

In the transverse-traceless gauge, $q^\mu h_{\mu\nu}=0$ and $h^\mu_{\mu}=0$, the graviton field $h_{\mu\nu}(q,z)$ satisfies the linearized Einstein equation \cite{Abidinr,AbN}:
\begin{equation}
z^3\partial_z\left(\frac{1}{z^3}\partial_z h_{\mu\nu}(q,z)\right) + q^2 h_{\mu\nu}(q,z)=0.
\label{hbulk}
\end{equation}

The bulk-to-boundary propagator $h(q,z)$ satisfies the normalization condition and follows the ultraviolet boundary $h(q, \epsilon) = 1$. The surface term from the IR boundary obtained when varying the action then vanished and solution to 
(\ref{hbulk})
\begin{equation}
h(q,z)=\frac{\pi}{4}q^2 z^2 \left[ \frac{Y_1(qz_0)}{J_1(qz_0)}J_2(qz) - Y_2(qz)\right].    
\end{equation}
For spacelike momentum transfer $q^2=-Q^2<0$. The corresponding bulk-to-boundary propagator is obtained as
\begin{equation}
\mathcal{H}(Q,z)=\frac{1}{2}Q^2 z^2 \left[ \frac{K_1(Qz_0)}{I_1(Qz_0)}I_2(Qz) + K_2(Qz)\right], \label{bulkpro}
\end{equation}
which smoothly interpolates between the boundary value $\mathcal{H}(Q,\epsilon)=1$ and the confined dynamics at $z=z_0$.

\section{Deuteron Gravitational Form Factors}

Traditionally, much of our understanding of hadronic structure arises from electromagnetic form factors (EFF), which probe charge and current distributions inside hadrons. Along with the EFFs, the GFFs help to take deeper insight into the internal dynamics of a hadron. Studies of the matrix elements of the EMT reveal how energy, momentum, and internal forces are distributed. The corresponding quantities, known as GFFs, provide access to the mass radius of the hadron, the distributions of pressure and shear forces, and the mechanical stability of strongly interacting systems.

The deuteron GFFs encode the internal distributions of energy and stress and are defined through the EMT matrix element between deuteron states. In the holographic framework, the EMT couples to the five-dimensional graviton field \(h_{\mu\nu}\), and the expectation value of the first follows from the variation of the 5D action:
\begin{equation}
T^{\mu\nu} = -\frac{2}{\sqrt{-g}} \frac{\delta S}{\delta h_{\mu\nu}}.
\label{4.1}
\end{equation}

Due to EMT symmetry and conservation, \(T^{\mu\nu}\) has six independent components, which decompose into a transverse-traceless part \(\hat{T}^{\mu\nu}\) and a trace part \(\tilde{T}^{\mu\nu}\):
\begin{equation}
T^{\mu\nu} = \hat{T}^{\mu\nu} + \tilde{T}^{\mu\nu},
\qquad 
\tilde{T}^{\mu\nu}=\frac{1}{3}\!\left(\eta^{\mu\nu}-\frac{q^{\mu}q^{\nu}}{q^2}\right)\!T.
\label{4.2}
\end{equation}
The physical contribution relevant to the deuteron arises from \(\hat{T}^{\mu\nu}\), while \(\tilde{T}^{\mu\nu}\) vanishes under the gauge constraints.

Within the AdS/QCD correspondence, the three-point function involving two deuteron vector currents and one EMT insertion is obtained from the cubic variation of the 5D action:
\begin{equation}
\langle 0 | \mathcal{T} J^\alpha(x)\, \hat{T}^{\mu\nu}(y)\, J^\beta(w) | 0 \rangle =
-\,\frac{2\,\delta^3 S}{\delta V^0_\alpha(x)\, \delta h^0_{\mu\nu}(y)\, \delta V^0_\beta(w)}.
\label{4.7}
\end{equation}

The EMT matrix element between deuteron states can be parameterized as
\begin{eqnarray}
\langle d^a(p_2,\lambda_2) | \hat{T}^{\mu\nu}(q) | d^b(p_1,\lambda_1) \rangle
&=& (2\pi)^4 \delta^{(4)}(q+p_1-p_2)\, \delta^{ab}\,
\varepsilon^{*}_{2\alpha} \varepsilon_{1\beta} \nonumber\\[1ex]
&&\times\Big[
- A(Q^2)\Big(4 q^{[\alpha}\eta^{\beta](\mu} p^{\nu)} + 2\eta^{\alpha\beta}p^\mu p^\nu \Big)
\nonumber\\
&&\quad
-\frac{1}{2}\hat{C}(Q^2)\eta^{\alpha\beta}\big(Q^2\eta^{\mu\nu}-q^\mu q^\nu\big)
\nonumber\\
&&\quad
+ D(Q^2)\big(Q^2\eta^{\alpha(\mu}\eta^{\nu)\beta}
-2q^{(\mu}\eta^{\nu)(\alpha}q^{\beta)}+\eta^{\mu\nu}q^\alpha q^\beta\big)
\nonumber\\
&&\quad
- \hat{F}(Q^2)\frac{q^\alpha q^\beta}{m_D^2}\big(Q^2\eta^{\mu\nu}-q^\mu q^\nu\big)
\Big],
\label{4.11}
\end{eqnarray}
where \(p=(p_1+p_2)/2\) and \(q=p_2-p_1\). The invariant GFFs \(A(Q^2)\), \(\hat{C}(Q^2)\), \(D(Q^2)\), and \(\hat{F}(Q^2)\) are determined holographically as
\begin{eqnarray}
A(Q^2) &=& Z_2(Q^2), \nonumber\\
\hat{C}(Q^2) &=& -\frac{1}{Q^2}\!\left(\frac{4}{3}Z_1(Q^2) - \Big(Q^2+\frac{8m^2}{3}\Big)Z_2(Q^2)\right), \nonumber\\
D(Q^2) &=& -\frac{2}{Q^2}Z_1(Q^2) + \!\left(1+\frac{2m^2}{Q^2}\right)Z_2(Q^2), \nonumber\\
\hat{F}(Q^2) &=& \frac{4m^2}{3Q^4}\big(Z_1(Q^2) - m^2 Z_2(Q^2)\big),
\label{4.12}
\end{eqnarray}
with structure functions
\begin{eqnarray}
Z_1(Q^2) &=& \int^{z_0}_0 \frac{dz}{z}\, \mathcal{H}(Q,z)\, \partial_z\psi_n(z)\,\partial_z\psi_n(z), \nonumber\\
Z_2(Q^2) &=& \int^{z_0}_0 \frac{dz}{z}\, \mathcal{H}(Q,z)\, \psi_n(z)\,\psi_n(z).
\label{zler}
\end{eqnarray}

The gravitational root-mean-square radius, which describes the energy density distribution inside the deuteron, is extracted from the slope of \(A(Q^2)\) at zero momentum transfer, and its value for the $n=1$ state can be found using (\ref{zler}):
\begin{equation}
\langle r^2 \rangle_{\mathrm{grav}} = -6\,\left.\frac{\partial A(Q^2)}{\partial Q^2}\right|_{Q^2=0} = 0.486~\mathrm{fm}^2.
\label{4.14}
\end{equation}
For the radius, this gives $r_{\mathrm{grav}}\simeq0.697~\mathrm{fm}$, and this value agrees with the value  $\sqrt{ \langle R_m^2 \rangle }=0.68\pm0.01~\mathrm{fm}$  obtained in \cite{Lin} using CLAS experiment data fitted to the Uniform model parameterization. Also, it agrees with the result of Global Average for the Uniform model using near threshold $\rho^0$ photoproduction data measured by the ABHHM Collaboration \cite{Lin, ABH}. near threshold $\phi$ photoproduction data measured by the CLAS Collaboration \cite{Lin, Mibe} and the LEPS Collaboration \cite{Lin, Chang, Mibe2}. The value (\ref{4.14}) is close to the final extracted deuteron mass radius obtained from different form factor models using the Uniform model.

 \begin{figure*}[htbp]
		\begin{minipage}[c]{0.98\textwidth}
			{a)}\includegraphics[width=7cm,clip]{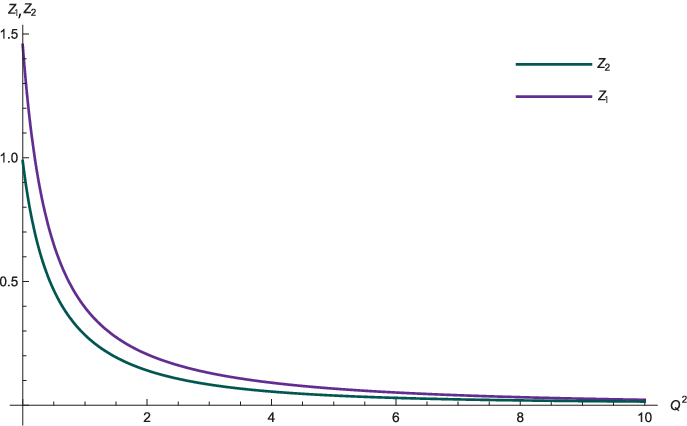}
		{b)}\includegraphics[width=7cm,clip]{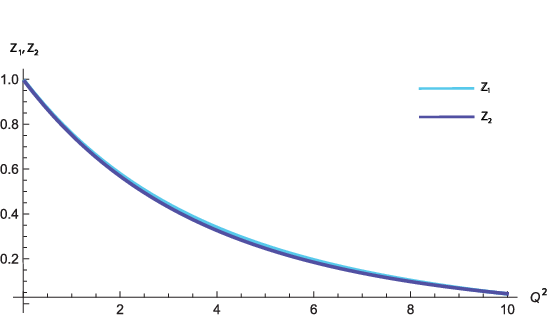}
		\end{minipage}
		\caption{Plot of $Z_1\left(Q^2\right)$ and $Z_2\left(Q^2\right)$ structure functions in AdS/QCD hard-wall a) and soft-wall b) models.}
		\label{gffplot}
	\end{figure*}	

	 \begin{figure*}[htbp]
		\begin{minipage}[c]{0.98\textwidth}
			{a)}\includegraphics[width=7cm,clip]{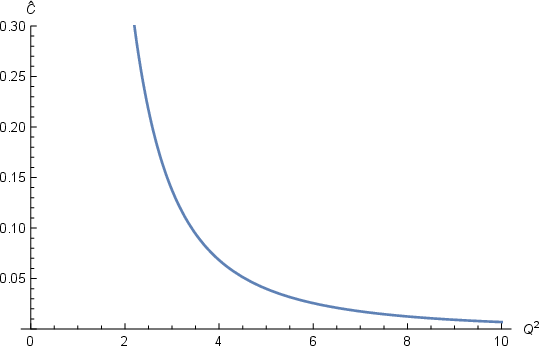}
			{b)}\includegraphics[width=7.5cm,clip]{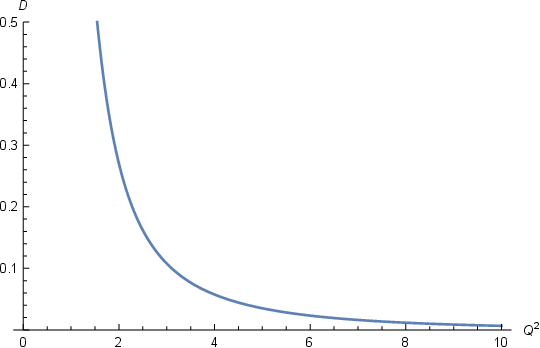}
            {c)}\includegraphics[width=7cm,clip]{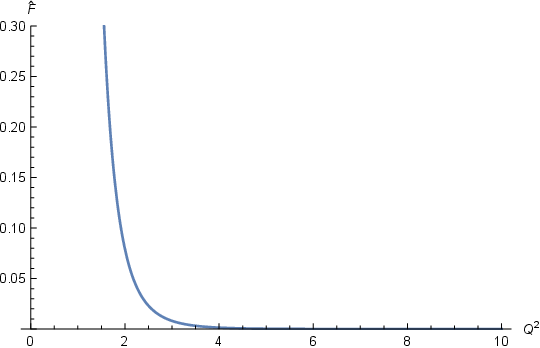}
		\end{minipage}
		\caption{Deuteron GFFs $\hat{C}(Q^2)$-a), $D(Q^2)$-b), $\hat{F}(Q^2)$-c) are in momentum space in the hard-wall model.}
		\label{gffplot2}
	\end{figure*}	

\label{sec:rho_b}

Fig.~\ref{gffplot} shows that deuteron gravitational form factor $A(Q^2)$, identified in this work with the form factor $Z_2(Q^2)$, as a function of momentum transfer $Q^2$ (in $~\mathrm{GeV^2}$). As seen in the graphs, in both models $Z_2(0)=1$, which means that the normalization condition $A(0)=1$ is satisfied, in agreement with the momentum sum rule for the energy–momentum tensor. With increasing $Q^2$, the form factor decreases monotonically and approaches zero, indicating a spatially localized mass–energy distribution inside the deuteron. To normalize $Z_1$, we divide this form factor by $Z_1(0)=1.24$.

Here, in Fig.~\ref{gffplot}, in the soft-wall model these FFs were plotted in $\mathrm{Gev^2}$ units, while in Refs.\cite{Mondal} these graphs were plotted in $\mathrm{fm^{-1}}$ units. Both graphs have the same behavior shape. In addition, comparing graphs a) and b) in Fig.~\ref{gffplot}, it can be observed that the dependence $Q^2$ of $Z_{1,2}$ FFs in the hard-wall model is the same as in the soft-wall model.

\section{Deuteron generalized parton distributions}

The GPDs provide a unified framework connecting form factors and parton distribution functions, thereby encoding both spatial and momentum information about the internal structure of the deuteron. For a spin-1 hadron such as the deuteron, five independent quark GPDs, $H_i(x,\xi,t)$ ($i=1,\dots,5$), appear in the light-front correlator of bilocal quark operators. They are defined through the non-diagonal light-cone matrix element
\begin{eqnarray}
&& \int \frac{p^+ dy^-}{2\pi} e^{ixp^+y^-} 
\left\langle p_2,\lambda_2 \right|  \bar q\!\left(-\frac{y}{2}\right) \gamma^+  
q\!\left(\frac{y}{2}\right)\left| p_1, \lambda_1 \right\rangle_{y^+=0, y_\perp = 0}
\nonumber\\
&& = -2(\varepsilon_2^*\!\cdot\!\varepsilon_1) p^+ H_1(x,\xi,t)
-\left( \varepsilon_1^+ \varepsilon_2^*\!\cdot\! q 
- {\varepsilon_2^+}^* \varepsilon_1\!\cdot\! q \right) H_2(x,\xi,t)
\nonumber\\
&& \quad + \frac{p^+}{m_d^2} q\!\cdot\!\varepsilon_1 \, q\!\cdot\!\varepsilon_2^*  H_3(x,\xi,t)
-\left( \varepsilon_1^+ \varepsilon_2^*\!\cdot\! q 
+ {\varepsilon_2^+}^* \varepsilon_1\!\cdot\! q \right) H_4(x,\xi,t)
\nonumber\\
&& \quad +\ 2 p^+ \left( \frac{m^2}{(p^+)^2} \varepsilon_1^+ {\varepsilon_2^+}^*
+ \frac{1}{3} (\varepsilon_2^*\!\cdot\!\varepsilon_1) \right) H_5(x,\xi,t),
\end{eqnarray}
where $x$ is the longitudinal momentum fraction carried by the active quark, $\xi$ is the skewness parameter ($\xi=-\frac{q^+}{2P^+}$), and $t=q^2$ is the momentum transfer.  

At the first moment in $x$, the GPDs are related to the gravitational form factors through the energy–momentum tensor (EMT) decomposition:
\begin{align}
\int_{-1}^1 x\, dx \, H_1(x,\xi,t) &=A(t) - \xi^2\hat{C}(t) + \frac{t}{6m^2} D(t),	\nonumber \\
\int_{-1}^1 x\, dx \, H_2(x,\xi,t) &= 2\left( A(t)+B(t) \right),	\nonumber \\
\int_{-1}^1 x\, dx \, H_3(x,\xi,t) &= E(t) + 4 \xi^2 \hat{F}(t),	\nonumber \\
\int_{-1}^1 x\, dx \, H_4(x,\xi,t) &= -2 \xi D(t),	\nonumber \\
\int_{-1}^1 x\, dx \, H_5(x,\xi,t) &=  \frac{t}{2m^2} D(t), \label{gpd}
\end{align}
where the GPDs for the valence quark are defined as $\int^{1}_{-1}H_v(x,\xi,t)dx=\int^1_0H_v(x,\xi,t)dx+\int^1_0H_v(-x,\xi,t)dx$.

For the calculation of deuteron GPDs, we need the integral representations for the functions $I_2(Qz)$ and $K_2(Qz)$, similar to those for $K_1$ and $I_1$ used in Refs.\cite{Lhard, Brodsky} for the calculations of nucleon GPDs. Following these works, the propagator (\ref{bulkpro}) can be written in terms of such a representation. Taking $\nu=2$ in the integral representation for $I_{\nu}$ given in \cite{mat} and replacing $z$ by $zQ$ we get the desired form for $I_2$:

\begin{equation}
I_2(Qz)=\frac{2}{3\pi}Q^2z^2\int^1_0 dx(1-x^2)^{\frac{3}{2}}\cosh{(xzQ)}. \label{modifie}
\end{equation}
For $K_2$, we apply a general representation for $K_{\nu}$ used in \cite{Lhard}. Taking $\nu=2$, replacing $z$ by $zQ$ and making the change of variable $t=Q^2\frac{1-x}{4x}$, we obtain the following integral representation for the function $K_2$:
\begin{equation}
 K_2(Qz)=\frac{2z^2}{Q^2}\int^1_0 dx\frac{x\exp\Big(\frac{-Q^2(1-x)}{4x}-\frac{z^2x}{1-x}\Big)}{(1-x)^3}.\label{modif3}    
\end{equation}
With these representations of $K_2$ and $I_2$ functions, we can write the bulk-to-boundary propagator $\mathcal{H}(Q,z)$ in terms of integral over $x$:
\begin{eqnarray}
    &\mathcal{H}(Q,z)&=\int^1_0 dx \left[ \frac{4K_1(Qz_0)}{3 \pi I_1(Qz_0)}Q^4z^4\left(1-x^2\right)^{\frac{3}{2}}\cosh{(xzQ)}+\frac{z^4x\exp\Big(\frac{-Q^2(1-x)}{4x}-\frac{z^2x}{1-x}\Big)}{(1-x)^3}\right] \nonumber\\
    &&=\int^1_0 dx \mathcal{\tilde{H}}(x,Q,z),\label{btb}
\end{eqnarray} 
 where, $\mathcal{\tilde{H}}(x,Q,z)$ denotes the expression in square brackets. Then the integration variable $x$ in (\ref{modifie}) and (\ref{modif3}) is identified with the momentum fraction in the GPD-CFF relations (\ref{gpd}).
 Representation of $\mathcal{H}$ (\ref{btb}) enables us to write the form factors $Z_{1,2}$ and then the form factors $A, D$ (for case $\xi=0$), in terms of the integral over $x$ from $\mathcal{H}$. Finally, for GPDs $H_1, H_2, H_5$ we find the following integral expressions:
\begin{eqnarray}
    &&H_1(x,0,Q^2)=\frac{1}{x}\int^{z_0}_0 \frac{dz}{z}\mathcal{\tilde{H}}(x,Q,z) \left[ \left(\frac{4}{3}-\frac{6Q^2}{m^2}\right)\psi_n(z) \psi_n(z) - \frac{3Q^2}{m^2}\partial \psi_n(z) \partial \psi_n(z)\right], \nonumber\\
  &&H_2(x,0,Q^2)=\frac{2}{x}\int^{z_0}_0 \frac{dz}{z}\mathcal{\tilde{H}}(x,Q,z)\psi_n(z)\psi_n(z), \nonumber\\
  &&H_5(x,0,Q^2)=\frac{2}{3xQ^2}\int^{z_0}_0 \frac{dz}{z}\mathcal{\tilde{H}}(x,Q,z) \left[\partial \psi_n(z) \partial \psi_n(z)-m^2 \psi_n(z)\psi_n(z)\right].\label{*}
\end{eqnarray}

Since the first moment of $H_{2}$ is related to the spin $J_z$ of the spin-1 particle  \cite{Abidinr, Ji}, it can be checked by direct numerical calculation that this GPD in (\ref{*}) obeys the normalization condition:
\begin{equation}
         \int^1_1xdxH_2(x,0,0)=2(J_z)_{max}=2.
\end{equation}

	 \begin{figure*}[htbp]
		\begin{minipage}[c]{0.98\textwidth}
			{a)}\includegraphics[width=6cm,clip]{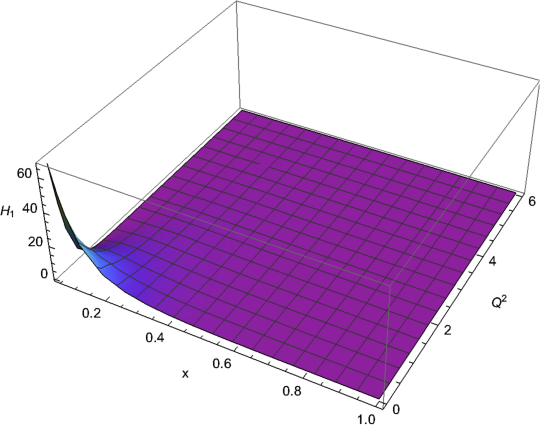}
			{b)}\includegraphics[width=6cm,clip]{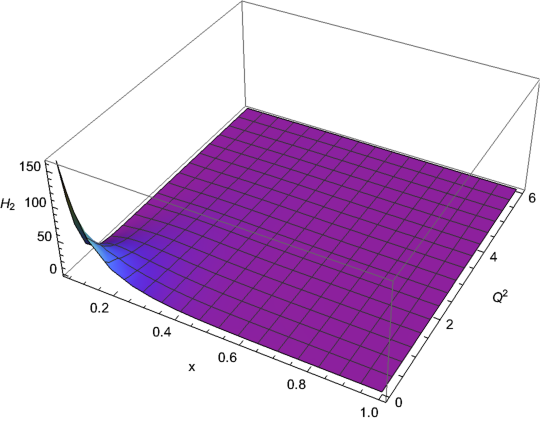}
            {c)}\includegraphics[width=6cm,clip]{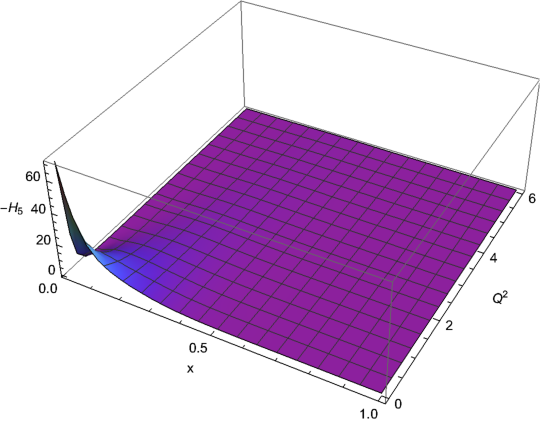}
		\end{minipage}
		\caption{Deuteron hard-wall GPDs $H_1$, $H_2$, and $H_5$ as a function of $x$ and $Q^2$ given in the a), b), and c) plots correspondingly.}
		\label{gpdplot}
	\end{figure*}	
    
    \begin{figure*}[htbp]
		\begin{minipage}[c]{0.98\textwidth}
			{a)}\includegraphics[width=6cm,clip]{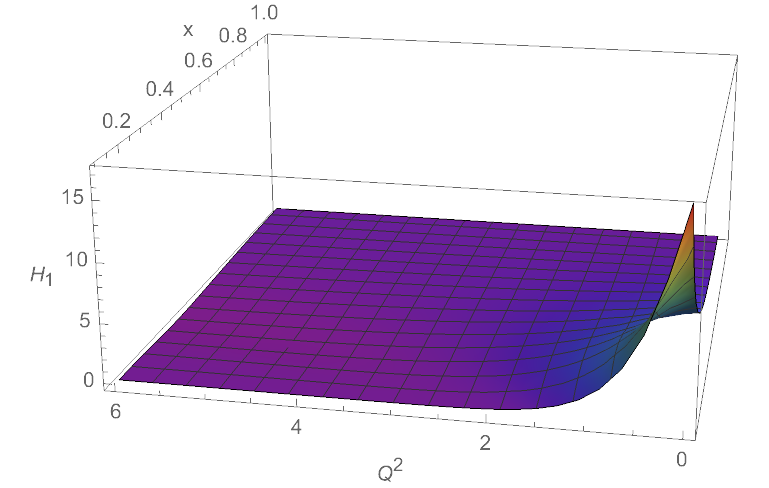}
			{b)}\includegraphics[width=6cm,clip]{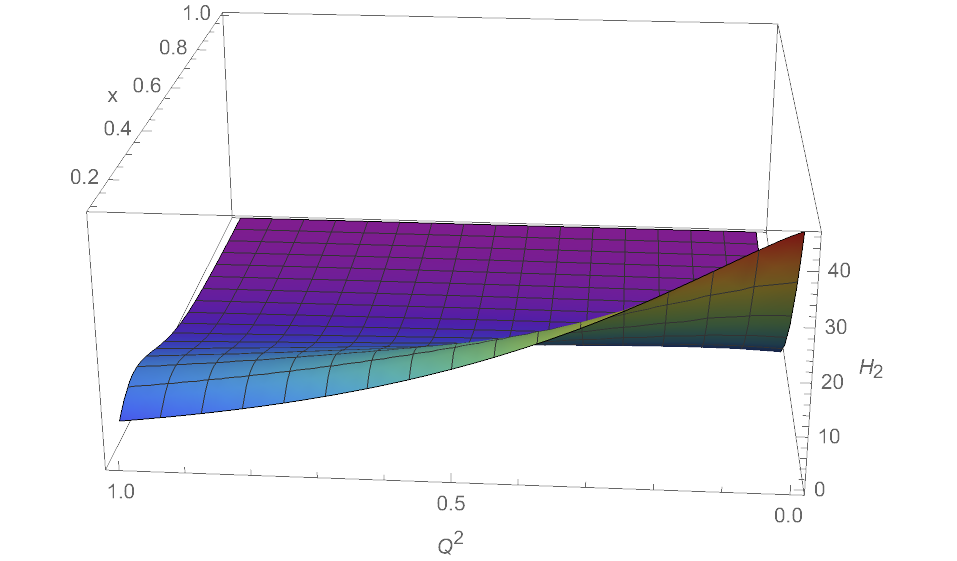}
            {c)}\includegraphics[width=6cm,clip]{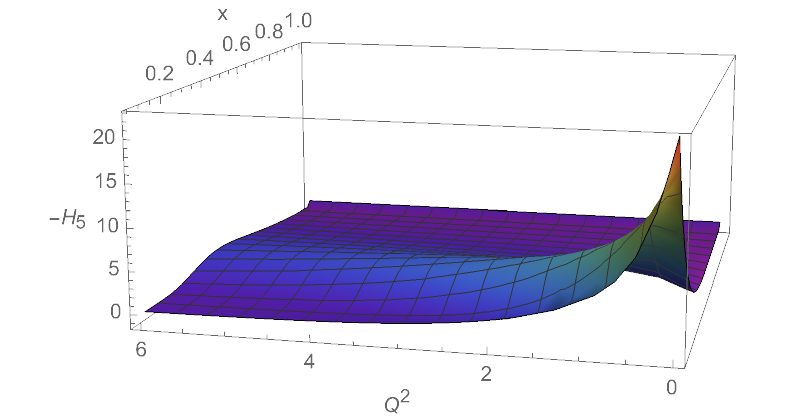}
		\end{minipage}
		\caption{Plots of the $\rho$ meson GPDs $H_i(x,0,Q^2)$ as a function of $x$ and $Q^2$ in the hard-wall model.}
		\label{gpdplot2}
	\end{figure*}	
Having explicit integral representations for GPDs $H_i$, we analyzed them graphically. The numerical results are shown in Fig. \ref{gpdplot}.
We can compare our graph for $H_1$ with the graph for GPD $H_v^1$ for a deuteron obtained in \cite{Mondal} from the EFFs within the soft-wall model (Fig.8 $a)$). We observe that the shapes of the GPD graphs plotted in Fig.\ref{gpdplot} are similar to the graph for $H_v^1$  GPD obtained in \cite{Mondal}, and in both figures the GPD graph increases when $Q^2$ increases.

The GPDs obtained from the GFFs for the $\rho$ mesons were defined in \cite{Abidinr}, and have the same expressions as for the deuteron (\ref{gpd}), differing only by profile functions. To additionally compare the deuteron GPDs with those for the $\rho$ mesons, we have plotted the GPD graphs for these mesons in Fig. \ref{gpdplot2} using the meson's profile in \cite{Abidinr} and the bulk-to-boundary propagator (\ref{btb}). For the calculation of the GPDs of the $\rho$ meson, the parameter $z_0$ was fixed at the known value $z_0=1/(322~MeV)$ \cite{3}. Comparing the graphs for the $H_1$ and $H_5$ GPDs, the graphs a) and c), respectively, in both figures \ref{gpdplot} and \ref{gpdplot2}, we observe a close similarity of the distributions for the deuteron with the GPDs for the $\rho$ meson. For the $H_2$ GPDs, the b) graphs in the figures show a more spread-out distribution for the meson, whereas for the deuteron, it is mostly localized at the center.

\section{Summary}

Our results from GFF calculations provide an estimate of the gravitational radius of the deuteron that qualitatively agrees with those obtained in lattice QCD and other phenomenological models.  The integral expressions for the GPDs, obtained from the GFFs via the sum rules, were derived.  The graph analysis of the GPDs shows that the hard-wall results are consistent with those derived within the lattice QCD and the phenomenological estimates.
Additionally, graph analysis of hard-wall GPDs of deuteron obtained from GFFs reveals that the distribution shape is similar to those obtained from soft-wall electromagnetic form factors for these particles \cite{Mondal}.
 This study demonstrates that holographic methods can provide valuable insight into the nonperturbative structure of nuclear bound states. Applying formulas for GPDs obtained here, the internal structure of the hybrid meson $J^{PC}=1^{-+}$ can also be studied~\cite{Loredana}.


\newpage
	
\begin{thebibliography}{99}


\bibitem{Muller} D. Müller, D. Robaschik, B. Geyer, F. M. Dittes, J. Horejsi, Fortsch. Phys. {\bf 42}, 101 (1994), [arXiv:9812448 [hep-ph]].
\bibitem{Ji3} X.-D. Ji, J. Phys. G {\bf 24}, 1181 (1998), [arxiv:9807358 [hep-ph]].
\bibitem{Ji} X. -D. Ji,  Phys. Rev. Lett.{\bf 78}, 610 (1997), [arxiv:9603249 [hep-ph]].
\bibitem{Pol2} M. V. Polyakov, P. Schweitzer, Int. J. Mod. Phys. A {\bf 33}, 1830025 (2018), [arxiv: 1805.06596[hep-ph]].
\bibitem{Bro} A.V. Belitsky, X. Ji, Phys. Lett. B {\bf 538}, 289 (2002), [arxiv:0203276 [hep-ph]]. 
\bibitem{Pol} M.~V.~Polyakov, Phys. \ Lett. \ B {\bf 555}, 57-62 (2003), [arXiv:hep-ph/0210165].
\bibitem{KMSh} N.~Kumar, C.~Mondal, N.~Sharma, Eur. \ Phys. \ J. \ A {\bf 53}, 237 (2017), [arXiv:1712.02110 [hep-ph]]. 
\bibitem{4} A. V. Radyushkin, "Nonforward parton distributions",Phys. Rev. D {\bf 56}, 5524 (1997), [arxiv: 9704207[hep-ph]].
\bibitem{expC} V. Yu. Alexakhin, Yu. Alexandrov, G. D. Alexeev, A. Amoroso, B. Badełek, F. Balestra, J. Ball, G. Baum, Y. Bedfer et al. (COMPASS Collaboration), Phys. Rev. Lett. {\bf 94}, 202002 (2005).
\bibitem{expA} A. DeGrush, A. Maschinot, T. Akdogan, R. Alarcon, W. Bertozzi, E. Booth, T. Botto, J.R. Calarco, B. Clasie et al. (BLAST Collaboration), Phys. Rev. Lett. {\bf 119}, 182501 (2017).
\bibitem{expF} F. Bradamante (on behalf of the COMPASS Collaboration), PoS {\bf 346}, 23rd International Spin Physics Symposium (2018), [arxiv: 1812.07281[hep-ex]].
\bibitem{Compass} COMPASS Collaboration, CERN-EP-2025-298, (2025) [arxiv: 2512.22311[hep-ex]].
\bibitem{Alice} ALICE Collaboration, Nature 648, 306-311 (2025), [arxiv: 2504.02393[nucl-ex]].
\bibitem{Aslan} Y. Guo, F. P. Aslan, X. Ji, M. G. Santiago, Phys. Rev. Lett. {\bf 135}, 261903 (2025), [arXiv:arXiv:2509.08037 [hep-ph]].
\bibitem{Dima} D. Watkins, D. Keller, "Direct Deep Neural-network Extraction of Generalized Parton Distributions", [arXiv:2512.21761 [hep-ph]].
\bibitem{Bel} A.V. Belitsky, D. Müller, A. Kirchner, Nucl. Phys. B {\bf 629}, 323-392 (2002), [arXiv:0112108 [hep-ph]].
\bibitem{Lat} Ph. Hägler, W. Schroers, J. Bratt, J. W. Negele, A. V. Pochinsky, R. G. Edwards, D. G. Richards, M. Engelhardt, G. T. Fleming et al. (LHPC Collaboration), Phys. Rev. D {\bf 77}, 094502 (2008).
\bibitem{Ber} E. R. Berger, F. Cano, M. Diehl, B. Pire, Phys. Rev. Lett.{\bf 87}, 142302 (2001), [arxiv:0106192 [hep-ph]].
\bibitem{Mal} J. M. Maldacena, Adv. Theor. Math. Phys. {\bf 2}, 231 (1998), [arXiv:9711200 [hep-th]].
\bibitem{Gubs} S. S. Gubser, I. R. Klebanov and A. M. Polyakov, Phys. Lett. B {\bf 428}, 105 (1998) [arXiv:9802109[hep-th]].
\bibitem{Wit} E. Witten, Adv. Theor. Math. Phys. 2, 253 (1998) [arXiv: 9802150[hep-th]].
\bibitem{Boschi1} H. Boschi-Filho and N.R.F. Braga, JHEP 0305, 009 (2003), [arXiv:0212207[hep-th]].
\bibitem{Boschi2} H. Boschi-Filho and N.R.F. Braga, Eur. Phys. J. C {\bf 32}, 529 (2004) [arXiv: 0209080[hep-th]].
\bibitem{Erl} J. Erlich, E. Katz, D. T. Son and M. A. Stephanov, Phys. Rev. Lett. {\bf 95}, 261602 (2005) [arxiv:0501128[hep-ph]]
\bibitem{Abidinr} Z.~Abidin, C.~E.~Carlson, Phys. \ Rev. D {\bf 77}, 095007 (2008), [arxiv:0801.3839 [hep-ph]].
\bibitem{3} Z.~Abidin, C.~E.~Carlson, Phys. \ Rev. D {\bf 78}, 071502 (2008), [arxiv:0808.3097 [hep-ph]].
 \bibitem{AbN}  Z. Abidin and C. E. Carlson, Phys.\ Rev. \ D, {\bf 79}, 115003 (2009), [arXiv:0903.4818 [hep-ph]].
 \bibitem{Lubsoft} A. Vega, I. Schmidt, T. Gutsche, V. E. Lyubovitskij, Phys. Rev. D {\bf 83}, 036001 (2011), [arXiv:hep-ph/1010.2815v3].
 \bibitem{Sharma} N. Sharma,  Phys. Rev. D {\bf 90}, 095024 (2014), [arXiv:hep-ph/1411.7486].
\bibitem{Mro} M.~Allahverdiyeva, S.~Mamedov, Eur. \ Phys. \ J. \ C {\bf 83}, 5, 447 (2023), [arXiv:2302.03383 [hep-ph]].
\bibitem{lyudeyt} T.~Gutsche, V.~E.~Lyubovitskij, I.~Schmidt, Phys. Rev. D {\bf 94}, 116006, (2016), [arXiv:1607.04124 [hep-ph]].
\bibitem{lyudeyt2} T.~Gutsche, V.~E.~Lyubovitskij, I.~Schmidt, A. Vega, Phys. Rev. D, {\bf 91}, 114001, (2015), [arxiv:1501.02738 [hep-ph]].
\bibitem{6} N.~Huseynova, Sh.~Mamedov, J.~Samadov, Chin. \ Phys. C {\bf 47}, 013104, (2023), [arxiv:2204.06205 [hep-ph]].
\bibitem{Mondal} C.~Mondal, D.~Chakrabarti, X.~Zhao, Eur.\ Phys.\ J.\ A {\bf 53}, 106 (2017), [arXiv:1705.05808 [hep-ph]].
\bibitem{Md} S.~Mamedov, M.~Allahverdiyeva and N.~Akbarova, Eur. \ Phys. \ J. \ C {\bf 85}, 361 (2025), [arXiv:2412.17407 [hep-ph]]. 
\bibitem{Shr} Sh. Mamedov, N. Akbarova, M. Allahverdiyeva, PEPAN {\bf 20}, 1477–1479, (2023), [arXiv:2305.09704 [hep-ph]].
\bibitem{Brod} S.J. Brodsky, G.F. de Teramond, Phys. Rev. D {\bf 78}, 025032 (2008) [arXiv:0809.4899 [hep-ph]].
\bibitem{Fujii} D.~Fujii, A.~Iwanaka, M.~Tanaka, Phys. Rev. D {\bf 110}, L091501 (2024), [arXiv: 2407.21113 [hep-ph]].
\bibitem{Fujii2} D.~Fujii, M.~Kawaguchi, M.~Tanaka, Phys. Lett.B {\bf 866} 139559 (2025), [arXiv:2503.09686 [hep-ph]].
\bibitem{Fujii3} D.~Fujii, A.~Iwanaka, M.~Tanaka, Phys. Rev. D {\bf 112}, 094051 (2025) [arXiv:2507.18690 [hep-ph].
\bibitem{Jaume} J.~Garriga and T.~Tanaka,  Phys. Rev. Lett. {\bf 84}, 2778 (2000), [arXiv:9911055 [hep-th]].	
\bibitem{motoi}  N.~Maru, M.~Tachibana, Eur. Phys. J. C {\bf 63},123 (2009), [arXiv:0904.3816 [hep-ph]].
\bibitem{Lin} X.~Lin, W.~Kou, S.~Fu, R.~Wang, C.~Han, X.~Chen, Chinese Physics C {\bf 49}, 10 (2025), [arxiv:2504.10023 [hep-ph]]. 
\bibitem{ABH} Benz, P. and Braun, O. and Butensch{\"o}n, H. and Gall, D. and Idschok, U. and Kiesling, C. and Knies, G. and M{\"u}ller, K. and Nellen, B. and Schiffer, R. and Schlamp, P. and Schnackers, H. J. and Spiering, P. and Stiewe, J. and Storim, F., Nucl. Phys. B {\bf 79}, 10 (1974).
\bibitem{Mibe}  T. Mibe et al. (CLAS Collaboration), Phys. Rev. C {\bf 76}, 052202 (2007), [arxiv:0703013 [nucl-ex]].
\bibitem{Chang} W. C. Chang et al.,  Phys. Lett. B {\bf 658}, 209 (2008), [arXiv:nucl-ex/0703034].
\bibitem{Mibe2}T. Mibe et al. (LEPS Collaboration), Phys. Rev. Lett. {\bf 95}, 182001 (2005) [arxiv:0506015 [nucl-ex]].
 \bibitem{Lhard} A.~Vega, I.~Schmidt, T.~Gutsche, V.~E.~Lyubovitskij, Phys. \ Rev. D {\bf 85}, 096004 (2012), [arxiv:1202.4806 [hep-ph]]. 
 \bibitem{Brodsky} S. J. Brodsky, G. F. de Teramond, Phys. Rev. D {\bf 77},056007 (2008), [arXiv:0707.3859 [hep-ph]].
\bibitem{mat}http://functions.wolfram.com/03.02.07.0004.01
\bibitem{Loredana} L.~Bellantuono, P.~Colangelo, F.~Giannuzzi, Eur. Phys. J. C {\bf 74}, 2830 (2014), [arXiv:1402.5308 [hep-ph]]. 


\end{thebibliography}
    \end{document}